\title{A scalable architecture for quantum computation with molecular nanomagnets}
\author{M. D. Jenkins,\textit{$^{a,b\ddag}$} D. Zueco,\textit{$^{a,b,c}$} O. Roubeau,\textit{$^{a,b}$} G. Arom\'{\i},\textit{$^{d}$}\\ J. Majer,\textit{$^{e}$} and F. Luis\textit{$^{a,b}$}
}
\date{August 22, 2016}

\documentclass[12pt]{article}
\usepackage{graphicx}
\usepackage[version=3]{mhchem}
\usepackage{siunitx}
\usepackage{a4wide}

\newcommand\blfootnote[1]{%
  \begingroup
  \renewcommand\thefootnote{}\footnote{#1}%
  \addtocounter{footnote}{-1}%
  \endgroup
}

\begin{document}
\maketitle

\begin{abstract}
A proposal for a magnetic quantum processor that consists of individual molecular spins coupled to superconducting coplanar resonators and transmission lines is carefully examined. We derive a simple magnetic quantum electrodynamics Hamiltonian to describe the underlying physics. It is shown that these hybrid devices can perform arbitrary operations on each spin qubit and induce tunable interactions between any pair of them. The combination of these two operations ensures that the processor can perform universal quantum computations. The feasibility of this proposal is critically discussed using the results of realistic calculations, based on parameters of existing devices and molecular qubits. These results show that the proposal is feasible, provided that molecules with sufficiently long coherence times can be developed and accurately integrated into specific areas of the device. This architecture has an enormous potential for scaling up quantum computation thanks to the microscopic nature of the individual constituents, the molecules, and the possibility of using their internal spin degrees of freedom.
\end{abstract}

\blfootnote{\textit{$^{a}$~Instituto de Ciencia de Materiales de Arag\'on, Universidad de Zaragoza, Zaragoza, Spain. Tel: 34 876 553342; E-mail: fluis@unizar.es}}
\blfootnote{\textit{$^{b}$~Departamento de F\'{\i}sica de la Materia Condensada, Universidad de Zaragoza, Zaragoza, Spain.}}
\blfootnote{\textit{$^{c}$~Fundacion ARAID, 50004 Zaragoza, Spain.}}
\blfootnote{\textit{$^{d}$~Departament de Qu\'{\i}mica Inorg\`{a}nica, Universitat de Barcelona, Barcelona, Spain.}}
\blfootnote{\textit{$^{e}$~Vienna Center for Quantum Science and Technology, Atominstitut, TU Wien, 1020 Vienna, Austria. }}

\blfootnote{\dag~Electronic Supplementary Information (ESI) available: . See DOI: }
%additional addresses can be cited as above using the lower-case letters, c, d, e... If all authors are from the same address, no letter is required

\blfootnote{\ddag~Present address: Kavli Institute of Nanoscience, Delft University of Technology, Delft, The Netherlands.}

\newpage

\section{Introduction}
\label{Introduction}

Quantum information \cite{Ladd2010,Nielsen2011} is not only one of the most dynamical and fascinating branches of science, it is also seen by many as the technological revolution of the $21^{\rm st}$ century. Quantum coherence and entanglement give resources to crack tough computational problems, relevant to the design of new chemicals and materials, the safe data protection and communication and the efficient search in large data bases, which are beyond those affordable by any classical device. An outstanding challenge, common to existing schemes based on either trapped ions or solid state devices, is to scale up quantum computation architectures to a level where they are of practical use in these applications.\cite{Schoelkopf2008}

\begin{figure}[ht]
 \centering
 \includegraphics[height=6cm]{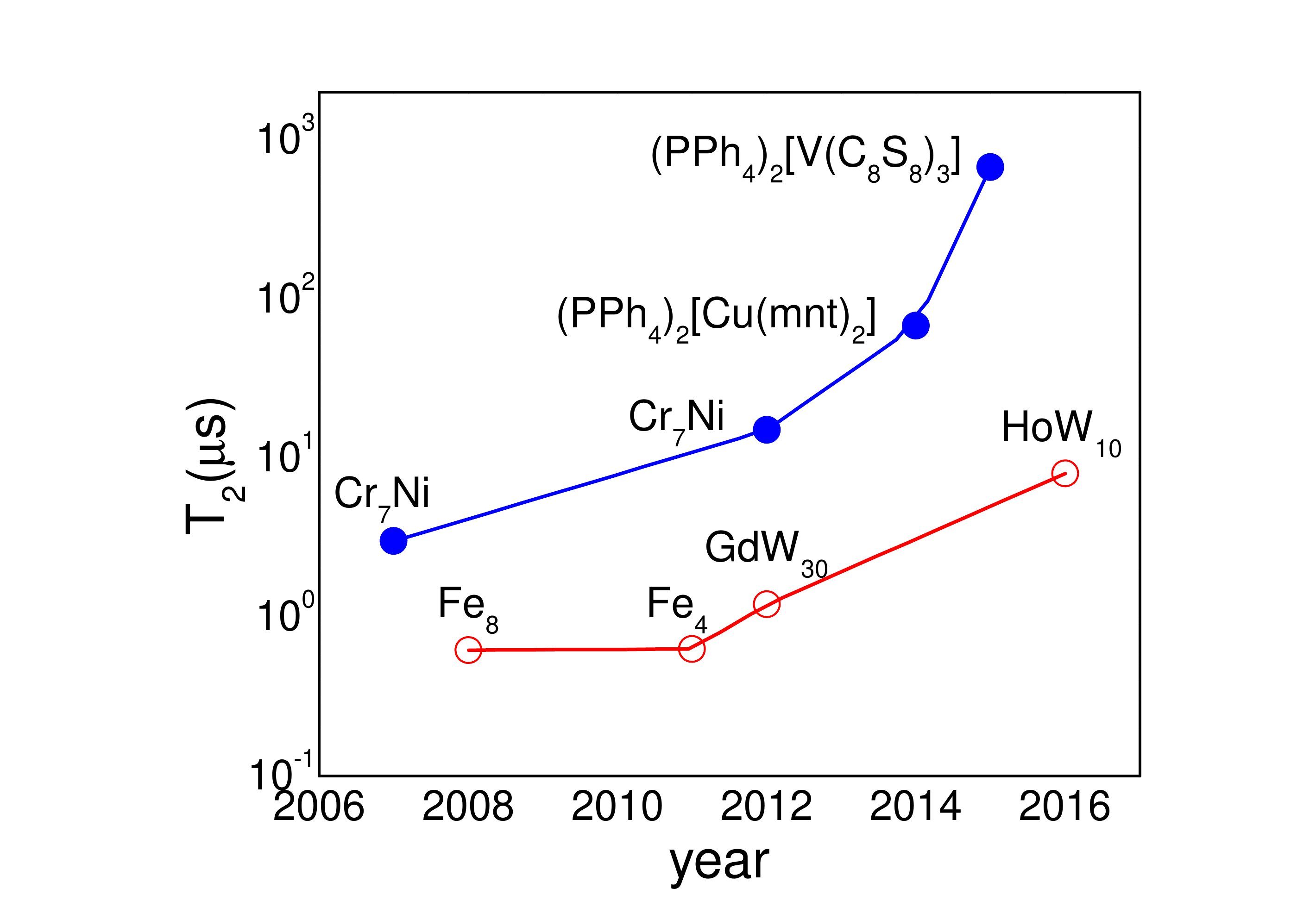}
 \caption{Recent progress in the spin coherence times of molecular nanomagnets with either $S=1/2$, $\bullet$, or $S> 1/2$, $\circ$. Data for the former are taken from Refs. \cite{Ardavan2007,Wedge2012,Bader2014,Zadrozny2015} whereas those for the latter correspond to Refs. \cite{Martinez-Perez2012,Schlegel2008,Takahashi2011,Shiddiq2016}.}
 \label{fgr:Coherence}
\end{figure}

Molecular nanomagnets\cite{Sessoli2006,Luis2014} consist of a magnetic core, containing one or several magnetic ions, which is surrounded and held together by organic ligands. They joined the list of quantum hardware candidates about a decade ago when it was shown that qubit states might be encoded using the different molecular spin orientations and their quantum superpositions.\cite{Leuenberger2001,Troiani2005,Ardavan2007} A particularly attractive feature is that macroscopic numbers of identical molecules can be synthesized by a single chemical reaction and that their magnetic properties, thus the relevant parameters that define the qubit frequency and states, are amenable to chemical design.\cite{Martinez-Perez2012,Baldovi2012,Shiddiq2016} Chemistry enables also the realization of rigid molecular structures with a low concentration of nuclear spins. This strategy has led to a spectacular progress, shown in Fig. \ref{fgr:Coherence}, in enhancing spin coherence times to maximum values close to ms.\cite{Wedge2012,Bader2014,Zadrozny2015,Bader2016,Atzori2016} Besides, isolated molecular qubits can be grafted to surfaces\cite{Gatteschi2009,Mannini2010,Cornia2011,Domingo2012} and to other nanostructures, like carbon nanotubes,\cite{Urdampilleta2011,Ganzhorn2013} and can also be integrated into nanoelectronic devices, such as nanocontacts prepared by electromigration.\cite{Heersche2006,Burzuri2012,Vincent2012,Perrin2015} This possibility has allowed detecting the reversal of a single molecular spin and reading-out and coherently manipulating its nuclear spin state, using either magnetic or electric rf fields.\cite{Vincent2012,Thiele2013,Thiele2014}

In spite of this progress, a clear technology able to build a scalable computation architecture with these materials is still missing. Here, we describe in detail a proposal for an all-magnetic quantum processor. For this, we critically examine the possibility of using superconducting circuits to read-out, control and communicate molecular spin qubits. Our calculations are based on state-of-the art parameters for existing molecules and circuit designs. The results show that the idea is realizable. Besides, we describe the main challenges and propose a preliminary road map to overcome them. One of the aims of this work is to set well-defined goals that can serve as a guide for the further development of this field.

The paper is organized as follows. In section \ref{Description}, the basic idea is presented. Adapting previous work on circuit QED to the case of molecular spin qubits, it is also discussed how the coupling of these qubits to the superconducting circuit allows the realization of basic quantum operations. This discussion also sets threshold values for the spin coherence time and the coupling of each spin to photons that are required to carry out these operations. Sections \ref{Feasibility} and \ref{Integration} describe whether the proposal is technically feasible, {\em i.e.}, whether these threshold values can be attained via the fabrication of suitable superconducting devices and a proper integration of molecular qubits onto predefined circuit areas. Section \ref{Scalability} discusses the intrinsic potential of this proposal in terms of density of quantum information that can be processed by a single chip and of possibilities for creative design. Section \ref{Summary} summarizes the main results, the challenges lying ahead for the development of this technology and how chemistry can contribute to achieve the crucial milestones.

\section{Architecture and basic operations}
\label{Description}
A quantum computation is implementing the coherent evolution of a set of information units, or qubits, from a well defined initial state, the input in computational language, to a final, or output, state, which must be measured. Therefore, we should think of ways of building physical devices able to carry out such unitary evolutions in a controlled manner. In the following, we introduce a solid-state architecture based on magnetic molecules coupled to superconducting circuits, and discuss how these hybrid devices can perform quantum operations.

\begin{figure}[ht]
 \centering
 \includegraphics[height=6.5cm]{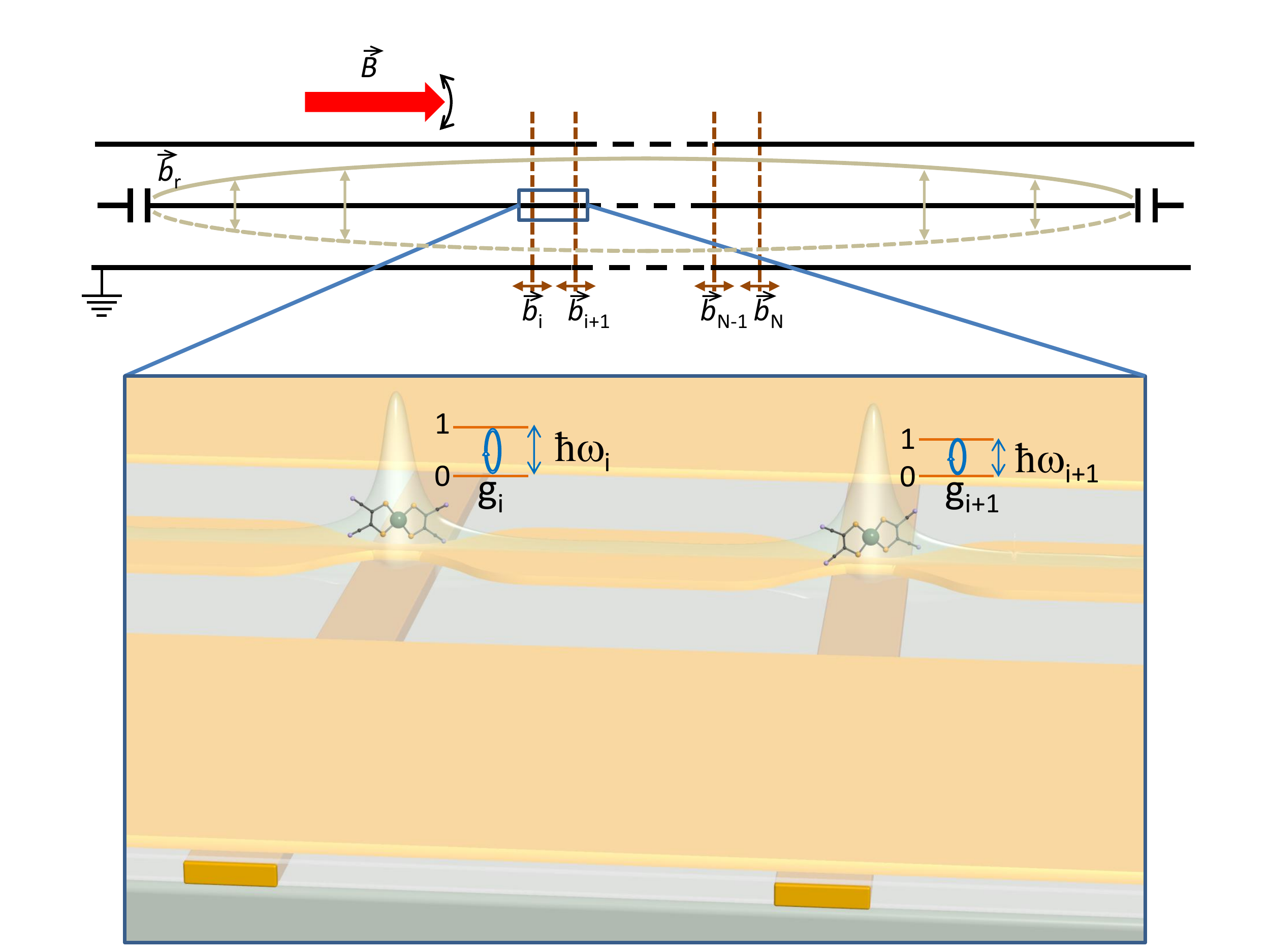}
 \caption{Top: Schematic image of a superconducting resonator and of the magnetic field profile $\vec{b}_{\rm r}$ of its ground $\lambda/2$ mode. An homogeneous in-plane magnetic field $\vec{B}$ and local magnetic fields $\vec{b}_{\rm i}$ generated by auxiliary lines (brown dotted lines) take the spin qubits in and out-of resonance with the resonator and induce single qubit operations. Bottom: Expanded artistic view of the central area of the magnetic quantum processor, showing that each molecular spin qubit rests near a nanoconstriction in the central resonator line, which enhances locally the microwave magnetic field, thus also the energy coupling $g_{\rm i}$ between each spin and a photon trapped in the resonator.}
 \label{fgr:MagQProcessor}
\end{figure}

\subsection{Overall description}
Any quantum operation can be decomposed as a set of single-qubit and two-qubit gates.\cite{Nielsen2011} A rather general strategy for scalability consists then of interconnecting a network of qubits via quantum channels which mediate the transfer of quantum information between nodes.\cite{Blais2004,Schoelkopf2008} This scheme, inspired by work on cavity quantum electrodynamics (QED), has been successfully implemented with solid-state superconducting devices: artificial atoms (solid-state qubits) couple to the electromagnetic field generated by a photon trapped in on-chip superconducting resonators.\cite{Wallraff2004} This strong coupling provides the opportunity to coherently control\cite{Schuster2005} and read-out\cite{Wallraff2005} the qubits, as well as to transfer information between different ones.\cite{Majer2007,DiCarlo2009,Niemczyk2010}

Large ensembles of spins, like NV$^{-}$ centers in diamond and others, have also been coherently coupled to such devices with the idea of using them as quantum memories.\cite{Schuster2010,Kubo2010,Wu2010,Amsuss2011} 
Concerning molecular systems, related proposals are to use the collective coupling between a molecular magnetic crystal and a resonator to define either a spin-photon hybrid qubit\cite{Carretta2013,Chiesa2016} or multiple qubits based on different spin-wave modes\cite{Wesenberg2009} In both approaches, quantum gates are performed by coupling to an auxiliary superconducting qubit, which provides the necessary nonlinear energy spectrum. An obvious alternative is to use the energy levels of individual molecules. It has been predicted that single molecular spins can show sufficiently strong couplings to quantum superconducting circuits, provided that suitable conditions are met.\cite{Jenkins2013} The use of single spins as qubits has also the advantage of minimizing the effect of dipole-dipole magnetic interactions, which constitute a major source of decoherence.\cite{Morello2006} Building on this idea, in this work we propose to apply circuit QED technology to read-out, coherently control, and interconnect individual molecular spin qubits.

A schematic view of the proposal is shown in Fig. \ref{fgr:MagQProcessor}. This magnetic quantum processor consists of three main components: a coplanar superconducting resonator, a set of individual magnetic molecules placed on specific locations of its central line, and a set of auxiliary superconducting wave guides perpendicular to the latter. The coplanar resonator consists of a central line coupled to the input and output leads by coupling capacitors and placed in between two quasi-infinite ground planes.\cite{Frunzio2005,Goppl2008} The chips are fabricated by depositing a thin film of a superconducting material (typically between $150$ and $300$ nm of Nb, Al, NbTi or even a high-$T_{\rm c}$ superconducting material such as YBaCuO \cite{Ghirri2015}) on a suitable substrate, like sapphire or silicon, and then using optical lithography to fabricate the lines and the coupling capacitors. These resonators support quantized electromagnetic photons with resonance frequencies $\omega_{\rm r}/2 \pi$ in the $1-10$ GHz region and really long lifetimes.\cite{Megrant2012,Bruno2015} Each magnetic molecule ${\rm i} = 1, N$ represents a qubit whose logic states $\mid 0 \rangle_{\rm{i}}$ and $\mid 1 \rangle_{\rm{i}}$ correspond to two mutually orthogonal magnetic energy states. The energy gap $\Delta_{\rm{i}}$ between the two levels associated with $\mid 0 \rangle_{\rm{i}}$ and $\mid 1 \rangle_{\rm{i}}$ can be tuned by an external homogeneous magnetic field $\vec{B}$ and by local fields $\vec{b}_{\rm i}$ generated by electrical currents flowing through the auxiliary lines. Depending on the orientation of $\vec{B}$, which determines the quantization axis of the qubits, these local fields can also induce transitions between the two qubit states. Each qubit couples also to the magnetic component $\vec{b}_{\rm r}$ of the resonator's electromagnetic field. In its fundamental mode, this component has nodes at the two resonator ends and a broad maximum at its center, where the molecules are to be placed. The coupling strength to a single photon trapped in the resonator is denoted by $g_{\rm i}$. The following sub-section provides a short description of the basic Hamiltonian that governs this hybrid system and that forms the basis for its quantum operation.

\subsection{Magnetic QED Hamiltonian}
\label{Hamiltonian}
The setup of Fig. \ref{fgr:MagQProcessor} can be described by the following Hamiltonian
\begin{equation}
\label{QED}
{\cal H} = \sum_{{\rm i}=1}^{N} {\cal H}_{\rm{mol,i}}
+
{\cal H}_{\rm r}
+
\sum_{{\rm i}=1}^{N} {\cal H}_{\rm{coupling,i}}
\end{equation}
\noindent The first term describes the magnetism of the isolated molecules and its response to external (and classical) magnetic fields, which together determine the qubit states $|0 \rangle_{\rm i}$ and $|1 \rangle_{\rm i}$ as well as the qubit energy gap $\hbar \omega_{\rm i}$. The second and third terms describe the quantized electromagnetic field in the resonator and its coupling to the spin qubits, respectively. In addition, one has to consider losses in the resonator, at a rate $\kappa$, and in the magnetic molecules, at a rate $\gamma$, respectively. In the former, losses are determined by the inverse of the quality factor $Q = \omega_{\rm r}/2 \pi \kappa$ (the number of coherent oscillations of an electromagnetic mode inside the resonator).\cite{Goppl2008,Megrant2012,Bruno2015} In the latter, they are determined by the decoherence of spin states, {\em e.g.} via the emission of phonons (rate ${\rm T}_{1}^{-1}$) or, mainly, by the couplings to nuclear spins that induce (at a rate ${\rm T}_{2}^{-1}$) phase shifts between different components of the spin wave function.\cite{Takahashi2011,Wedge2012,Bader2014} Magnetic dipolar interactions between molecules, which can dominate decoherence in ensembles,\cite{Morello2006} are expected to play almost no role, as different qubits are located very far apart in this scheme.

In the simplest scenario, when only second order anisotropy terms are relevant, the spin Hamiltonian of each molecule reads as follows: ${\cal H}_{\rm{mol,i}} = DS_z^2 + E (S_x^2 -S_y^2) -\mu_{\rm B}\vec{B}_{\rm i}\hat{g}_{S}\vec{S}$, with $\vec S$ the spin operators referred to principal anisotropy axes $x$, $y$ and $z$, $D$ and $E$ second order anisotropy constants, $\hat{g}_{S}$ the gyromagnetic tensor and $\vec{B}_{\rm i}$ the local magnetic field. In our proposal, the field has two components: an homogeneous magnetic field $\vec{B}$, applied by an external source (a magnet), and a local magnetic field $\vec{b}_{\rm i}$ generated by the auxiliary lines [Cf. Fig. \ref{fgr:MagQProcessor}]. The latter can have a dc and an oscillating component, {\rm i.e.} $\vec{b}_{\rm i} = \vec{b}_{\rm{i,dc}}+\vec{b}_{\rm{i,ac}} \cos(\omega t)$. Since these are open transmission lines, the frequency $\omega$ can vary between typically $1$ and $10$ GHz.\cite{Clauss2013}

For molecules with a net spin $S=1/2$, such as the Cr$_{7}$Ni rings and mononuclear Cu(II) and V(IV) complexes,\cite{Troiani2005,Ardavan2007,Wedge2012,Bader2014,Zadrozny2015,Bader2016,Atzori2016} the qubit basis is formed by 'spin-up' and 'spin-down' projections along $\vec{B}_{\rm i}$. The magnetic field intensity and the effective gyromagnetic ratio $g_{S}$, which depends on the relative orientation of $\vec{B}_{\rm i}$ with respect to the molecular axes, determine the qubit frequency $\hbar \omega_{\rm i} = \mu_{\rm B}g_{S}B_{\rm i}$, with $g_{S} \simeq 2$. In the case of high-spin ($S>1/2$) molecules, two suitable definitions exist for the computational basis.\cite{Jenkins2013} The first one is to identify the logic states with two spin projections $|m \rangle$ along $z$, whose energies are split by the magnetic anisotropy, that is, $|0 \rangle_{\rm i} \simeq |+S \rangle_{\rm i}$ and $|1 \rangle_{\rm i} \simeq |+S-1 \rangle_{\rm i}$ for $D < 0$ and $|0 \rangle_{\rm i} \simeq |0 \rangle_{\rm i}$ and $|1 \rangle_{\rm i} \simeq |+1 \rangle_{\rm i}$ for $D >0$. A second natural choice is to use the two lowest-lying eigenstates of ${\cal H}_{\rm{mol,i}}$. In this case, off-diagonal anisotropy terms can give rise to a finite tunnel splitting even at zero field. In both cases, the magnetic field dependence of the qubit level splitting $\hbar \omega_{\rm i} \equiv \langle 1 | {\cal H}_{\rm{mol,i}} | 1 \rangle -\langle 0 | {\cal H}_{\rm{mol,i}} | 0 \rangle$ can be approximately written as  $\hbar\omega_{\rm i} \simeq \hbar\omega_{\rm i}(B_{\rm i}=0)+g_{S}\mu_{\rm B}B_{\rm i}$ where $g_{S}$ is again an effective gyromagnetic ratio.

In order to simplify the discussion, we shall consider in the analysis that follows a simplified version of the Hamiltonian (\ref{QED}) which is derived by projecting the original one onto a basis formed by the two logic states of each molecule. The magnetic QED Hamiltonian then reads as follows
\begin{eqnarray}
\label{QED2}
\cal{H} &=&  \sum_{{\rm i}=1}^{N} \left[\hbar \omega_{\rm i} \sigma_{z,\rm{i}}-\frac{g_{S}\mu_{\rm B}}{2}\vec{\sigma}_{\rm i}\vec{b}_{\rm{i,ac}} \cos(\omega t) \right]
+
\hbar \omega_{\rm r} a^\dagger a \\ \nonumber
&+&
\sum_{{\rm i}=1}^{N}g_{\rm i} \hat \sigma_{x,\rm{i}} (a^\dagger + a ),
\end{eqnarray}
\noindent where the first, second and third terms describe, respectively, the ensemble of spin qubits, coupled to magnetic fields $\vec{B}$ and $\vec{b}_{\rm i}$, the resonator and their mutual interaction. Here, $\sigma_{\alpha,\rm{i}}$ are Pauli matrices along the local qubit axes and $a$ and $a^{\dagger}$ are, respectively, annihilation and creation operators of photons in the resonator. For $B_{\rm i} \neq 0$, the qubit axes do not necessarily coincide with the local anisotropy axes of the molecule. The resonance frequency $\omega_{\rm r}$ of the coplanar resonator, typically of few GHz, can be easily adjusted by design to adapt it to the range of molecular transitions. Each of the molecules can be tuned in and out of resonance with the circuit by the local magnetic fields $b_{\rm{i,dc}}$ (further details on this are given in section \ref{Tuning} below). A crucial parameter, for the present purposes, is the coupling strength of the spins to the resonator quantized magnetic field $b_{\rm r} \sim (a + a^\dagger)$. It is given by\cite{Jenkins2013}
\begin{equation}
g_{\rm i} = \frac{g_{S} \mu_{\rm B}}{\sqrt{2}}\left| \langle 0| \vec{b}_{\rm r} (\vec r_{\rm i})\vec{S} |1 \rangle \right| \; .
\label{g}
\end{equation}
\noindent Its actual value is discussed in the next section \ref{Coupling} for different circuit designs and potential molecular qubits. In the rest of this section we show that Eq. (\ref{QED2}) is sufficient for performing universal quantum computation.

\subsection{Elemental quantum operations on molecular spin qubits}
\label{Operation}
All operations described below are carried out in the dispersive regime. This regime corresponds to a situation in which the qubits are detuned from the resonator, thus avoiding any transfer of energy excitations between both subsystems. It is then appropriate to define a frequency mismatch parameter $\Delta_{\rm i} \equiv \hbar (\omega_{\rm i} - \omega_{\rm r})$. The dispersive regime is defined by the condition $\Delta_{\rm i} > g_{\rm i}$ or, equivalently, $g_{\rm i}/\Delta_{\rm i} < 1$.
\subsubsection{Qubit initialization}
Each spin qubit naturally relaxes, at a rate $T_{1}^{-1}$, towards its ground state as temperature decreases. Initialization can then be achieved by operating the device at temperatures such that $k_{\rm B} T \ll  \hbar \omega_{\rm i}$. For typical values of the qubit frequencies in the range of $1-10$ GHz, a ground state population above $0.999$ is achieved for temperatures ranging from $7$ to $70$ mK.

\subsubsection{Operations on single qubits}
As said above, any computation can be decomposed into one and two qubit operations. Single qubit rotations, {\em i.e.}, transitions between any two superpositions of $|0\rangle_{\rm i}$ and $|1\rangle_{\rm i}$ for each molecule, can be induced by using magnetic field pulses generated by the auxiliary lines. A first method, which somehow mimics that used with superconducting qubits,\cite{Blais2004,Wallraff2005} is to tune $\omega_{\rm i}$ locally by a dc magnetic field $\vec{b}_{\rm{dc,i}}$ and then manipulate the spin states with microwave pulses applied through the resonator. Another possibility is available when $\vec{b}_{\rm{ac,i}}$ is not parallel to the qubit quantization axis $z$. A microwave pulse $\vec{b}_{\rm{ac,i}}\cos(\omega t)$ having $\omega = \omega_{\rm i}$ is then able to induce a transition between the two qubit states. The final state can be controlled by suitably choosing the pulse duration.
\subsubsection{Two-qubit operations}
Two-qubit gates are more difficult to implement. It is the challenge of controlling molecule-molecule interactions that largely justifies the architecture proposed here. The figure of merit is the turn on / off ratio of the interaction that must be tuned in situ in order to carry out each of the gates set by the different steps of a given algorithm. To see how to implement these interactions, we focus here onto the case of two molecules, ${\rm i}$ and ${\rm j}$, coupled to a resonator. Since molecule-molecule interactions are mediated by the resonator, we expect that taking the former out of resonance with the latter must tend to suppress any cross talk among them. This guess is confirmed by calculations. It can be shown that, for $g_{\rm i}/\Delta_{\rm i} < 1$, the resonator mediated interaction between the two molecules reads\cite{Zueco2009}
\begin{equation}
\label{two-qubit}
{\cal H}_{\rm{i,j}}
=
g_{\rm i} g_{\rm j} \left[
\frac{1}{\Delta_{\rm i}} + \frac{1}{\Delta_{\rm j}} - \frac{1}{\hbar (\omega_{\rm i} + \omega_{\rm r})}-  \frac{1}{\hbar(\omega_{\rm j} + \omega_{\rm r})}
\right] \sigma_{x, \rm{i}} \, \sigma_{x, \rm{j}}
\end{equation}
When the two qubits are in resonance with each other, that is, when $\Delta_{\rm i} = \Delta_{\rm j} \equiv \Delta$, this effective interaction induces a coherent evolution of their spin states at a frequency $\simeq g_{\rm i}g_{\rm j}/\Delta$. Such effective coupling of two qubits via a resonator (quantum bus) has been first proposed\cite{Blais2004} and then realized\cite{Majer2007} with superconducting qubits. Two-qubit gates can be implemented by controlling the time interval in which the interaction is active. The interaction can be effectively switched-on and off, as required by the gate operation, by detuning the two qubits from each other. It is worth mentioning again that, even when the interaction is on, the molecules are energetically detuned from the resonator. Therefore, the gate operation does not involve any energy exchange between these two systems.

\subsubsection{Qubit read-out}
Finally, we mention how to perform the read out of each qubit. The possibility of doing non-demolition measurements of the qubit state is based on the fact that, in the dispersive regime $g_{\rm i}/\Delta_{\rm i} < 1$, the energy level spacing of the coupled qubit-resonator system depends on the state of the qubit. The resonance frequency, which can be determined by measuring the transmission through the device, is then shifted by $-g_{\rm i}^{2}/\Delta_{\rm i}$ ($+g_{\rm i}^{2}/\Delta_{\rm i}$) when the qubit $i$ is in state $|0 \rangle_{\rm i}$ ($|1 \rangle_{\rm i}$). As with the previous operations, this idea has been put in practice with superconducting qubits\cite{Blais2004,Schuster2005} Different qubits can be read-out by tuning their respective energy mismatch parameters $\Delta_{\rm i}$, {\rm e.g.} by making all $\Delta_{\rm j}$, with j$\neq$i, much larger than $\Delta_{\rm i}$. Since qubit flips by the driving field are suppressed in either case, this allows probing the states of the qubits by monitoring the cavity transmission without altering them.

\section{Is it feasible?}
\label{Feasibility}

Whether the device operation outlined in the previous section is technically feasible depends mainly on making $g^{2}_{\rm i}/\Delta_{\rm i}$ sufficiently large with respect to dissipation, {\em i.e.} with respect to both $\kappa$ and ${\rm T}_{2}^{-1}$. This energy scale determines the rate at which two qubit gates operate (see Eq. (\ref{two-qubit})) and the ability to read-out the qubit state. The above condition is then required to ensure that gate operations are not disturbed by decoherence and that resonance peaks associated with qubit states $|0 \rangle_{\rm i}$ and $|1 \rangle_{\rm i}$ can be resolved experimentally. Since $g_{\rm i}/\Delta_{\rm i} < 1$ in the dispersive regime, this condition implies that the coupling $g_{\rm i}$ must be larger than both $\kappa$ and ${\rm T}_{2}^{-1}$. Achieving this {\em strong coupling limit}, defined by the condition $g_{\rm i} {\rm T}_{2}/h > 1$, for individual molecular spins represents a daunting challenge. Besides, it is necessary to tune the energies of the qubits in order to switch-on and off the resonator mediated couplings between them. These two technical requirements are discussed quantitatively in the two subsections that follow next.

\begin{figure*}
\centering
\includegraphics[width=\textwidth]{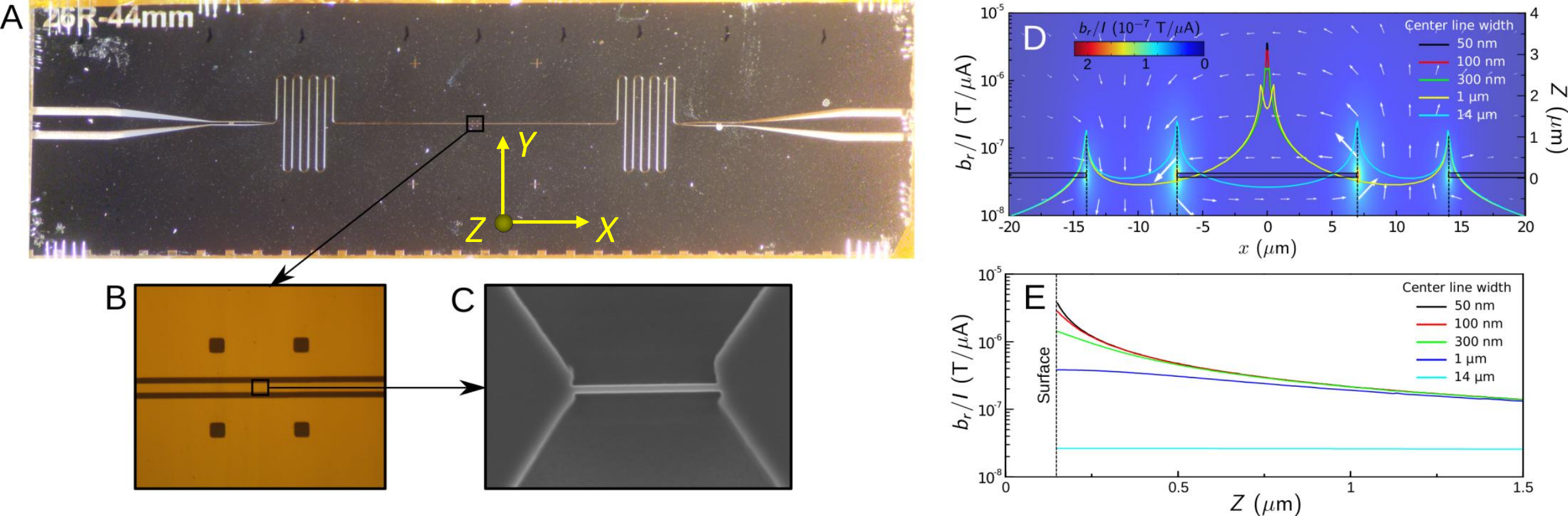}
\caption{A. Image of a coplanar superconducting resonator fabricated of Nb deposited onto saphire. For the ground, $\lambda/2$, cavity mode the radiation magnetic field shows a maximum in the central region, shown in B. By reducing the width of the central line in this region (panel C), the magnetic field intensity can be enhanced. D and E show, respectively, the magnetic field at the surface of the device as a function of $X$ (perpendicular to the central line) and at $X=0$ as a function of $Z$, the vertical distance above the substrate, for different central line widths $w$.  Panel D shows also, in the background, a contour plot of the magnetic field generated by the resonator in the $Y-Z$ plane.}
\label{fgr:Constriction}
\end{figure*}

\subsection{Spin-photon coupling and decoherence}
\label{Coupling}
The concept of circuit QED and the technology associated with it can be extended to diverse qubit realizations, provided that the energy coupling between qubits and photons is made sufficiently large as compared with the rates of decoherence. In the case of superconducting qubits, the large electric or magnetic dipolar moments make this coupling exceptionally strong.\cite{Wallraff2004,Niemczyk2010} For a single $S=1/2$ electronic spin, the typical coupling to a conventional resonator with a \SI{15}{\micro\meter} wide central line is of order $12$ Hz.\cite{Jenkins2013} In spite of the rather spectacular progress achieved in the last few years in enhancing spin coherence times (see Fig. \ref{fgr:Coherence}) this value corresponds to $g_{\rm i} {\rm T}_{2} < 8 \times 10^{-3}$, thus very far from the strong coupling regime. Here, we discuss how to locally enhance $g_{\rm i}$ via modifications of the circuit design. A closely related question is how to integrate the molecular spin qubits into these regions. This is left for a separate section \ref{Integration}.

The basic idea is illustrated in Fig. \ref{fgr:Constriction}, which shows an example of a Nb coplanar resonator. In its ground $\lambda /2$ mode, the  amplitude $b_{\rm r}$ of the microwave magnetic field vanishes at the two coupling capacitors, which mark the two ends of the cavity, and becomes maximum near the middle of the central line (the area shown in Fig. \ref{fgr:Constriction}B). This amplitude varies along the two directions, $Y$ (in plane) and $Z$ (vertical), perpendicular to the central line, showing sharp maxima near the edges of this line (Fig. \ref{fgr:Constriction}D) and decaying as one moves vertically from the surface (Fig. \ref{fgr:Constriction}E). The sharp maxima in $b_{\rm r} (Y)$ originate from the fact that superconducting currents flow mainly via a thin layer, of the order of the penetration depth, near the surface of the wire. If the width $w$ of the central line is made smaller, down to a few nm, the two peaks eventually merge into one giving rise to a large enhancement of the maximum $b_{\rm r}$. This effect can be seen in Figs. \ref{fgr:Constriction}D and E, which show the results of numerical simulations of $b_{\rm r}$ for resonators having constrictions of different widths. It has recently been shown that such nanoconstrictions can be fabricated by means of ion-beam nanolithography, that its presence does not affect much the resonance frequency and the intrinsic quality factor of the resonator, provided they are sufficiently short, say, $< \SI{1}{\micro\meter}$,\cite{Jenkins2014} and that these properties remain stable under magnetic fields up to $\sim 0.15$ T.\cite{Jenkins2015} A SEM image of a representative example is shown in Fig. \ref{fgr:Constriction}C.

\begin{figure}[ht]
\centering
\includegraphics[height=6cm]{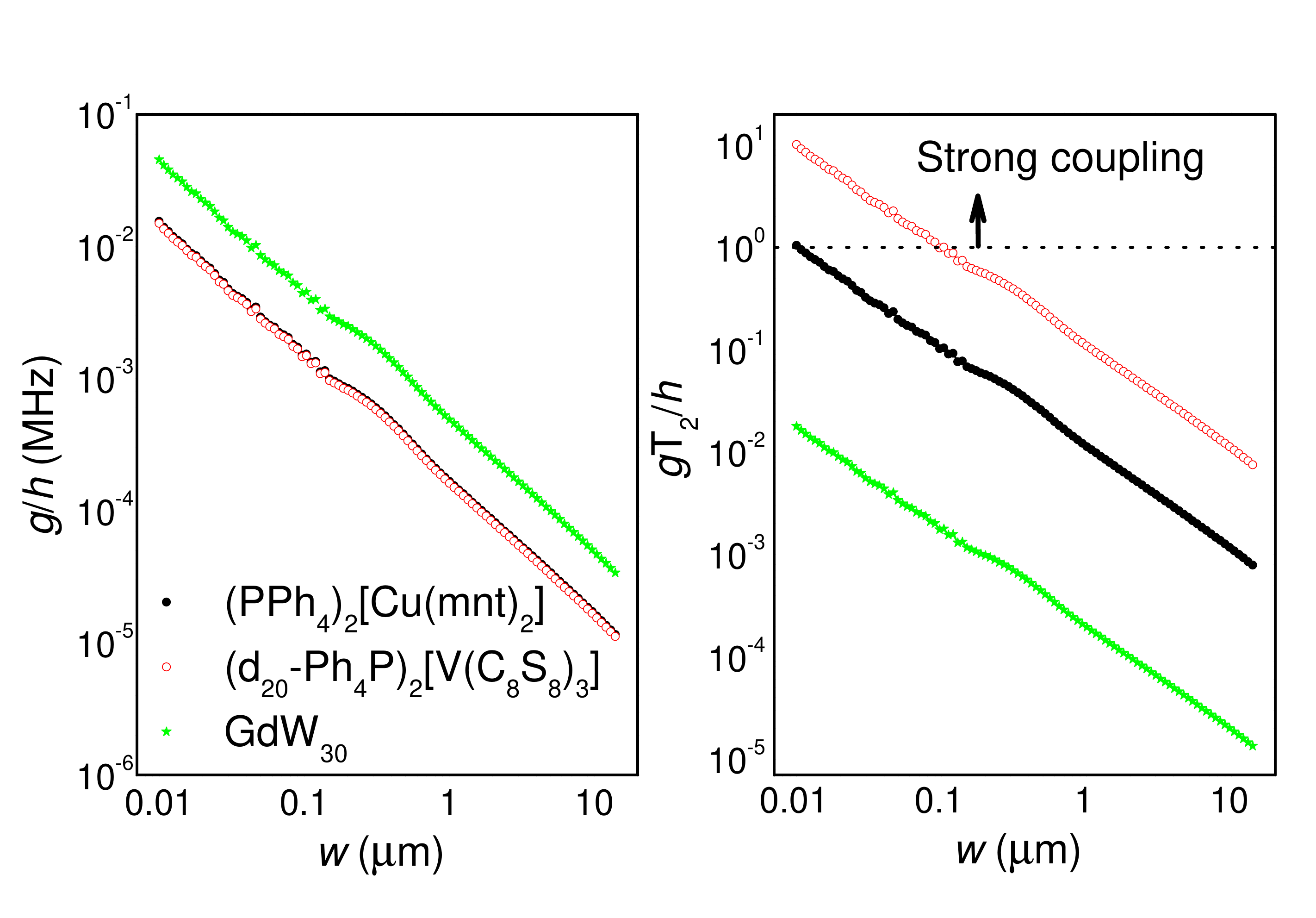}
\caption{Left: dependence of the single spin to single photon coupling $g$ on the width $w$ of the resonator central line calculated for different molecular spin qubits. In all calculations, the resonator and qubit frequencies were set to $6$ GHz, which corresponds to a magnetic field $B=0.43$ T for (PPh$_{4}$)$_{2}$[Cu (mnt)$_{2}$] and (d$_{20}$-Ph$_{4}$P)$_{2}$[V(C$_{8}$S$_{8}$)$_{3}$], both with $S=1/2$, and to $B \simeq 0$ for GdW$_{30}$. Right: same data multiplied by the low-temperature spin coherence times ${\rm T}_{2}$ of these molecules. The threshold for strong coupling, or coherent regime, is shown.}
\label{fgr:Coupling}
\end{figure}

The enhancement of the microwave field provides an opportunity to enhance also the coupling to magnetic molecules located at or near the constriction.\cite{Jenkins2013} Here, the small size of the molecular spin qubits can be seen as an advantage, provided that they can be integrated with sufficient accuracy. Figure \ref{fgr:Coupling} shows how the coupling of a $6$ GHz resonator to some qubit candidates depends on $w$. In these calculations, the molecules are located right on the center of the line ($Y=0$ and $Z=0$). The characteristic coupling strength shows a close to linear relation with $1/w$, increasing by three orders of magnitude as $w$ decreases from $\SI{14}{\micro\meter}$ down to $10$ nm. Preliminary experiments performed on free radical molecules coupled to $100$ nm wide constrictions confirm that the single spin coupling constant $g$ can be enhanced by more than two orders of magnitude with respect to that measured using conventional resonators.\cite{Jenkins2015}

For very narrow constrictions, the strong coupling limit can therefore be attained provided that coherence times are also sufficiently long. For instance, in the case of the (PPh$_{4}$)$_{2}$[Cu (mnt)$_{2}$] complex, with a low-temperature ${\rm T}_{2} \simeq 68 \mu$s,\cite{Bader2014} reaching this limit requires decreasing $w$ down to $10$ nm, which is close to the limit of nano-lithography technologies. The best situation is encountered for the nuclear-spin free (d$_{20}$-Ph$_{4}$P)$_{2}$[V(C$_{8}$S$_{8}$)$_{3}$],\cite{Zadrozny2015} also with a net $S=1/2$, which thanks to its record ${\rm T}_{2} \simeq 700 \mu$s might attain $g {\rm T}_{2} \simeq 10$. Reaching the strong coupling regime for $S=1/2$ molecules can, however, be also limited by the decoherence rate $\kappa$ of the circuits.

A way of further enhancing the coupling is to look for molecules with a spin $S>1/2$, such as lanthanide single-ion magnets.\cite{Layfield2015} However, the best ${\rm T}_{2}$ values reported to date for these qubit candidates are still rather modest (see Fig. \ref{fgr:Coherence}), and in most cases insufficient to reach strong coupling, as can be seen in Fig. \ref{fgr:Coupling}, which shows calculations performed for a GdW$_{30}$ polyoxometalate molecule having $T_{2} \simeq 1.2 \mu$s at low temperatures.\cite{Martinez-Perez2012} A promising possibility is to use tunnel split $\mid \pm m\rangle$ magnetic states to define the qubit basis.\cite{Jenkins2013} Clock transitions between these states have been shown\cite{Shiddiq2016} to be robust against decoherence induced by fluctuations in the local magnetic field and they can give rise to an enhancement of $g_{\rm i}$ by a factor $2m$ with respect to the simple case of a $S=1/2$ spin. For the recently studied HoW$_{10}$ polyoxometalate molecule, with $m=4$, attaining this goal requires that ${\rm T}_{2} > 8 \mu$s, which seems to be within reach.\cite{Shiddiq2016} However, because of the strong hyperfine coupling of Ho these states are excited states, thus they cannot be initialized by cooling. Finding similar phenomena in systems with weaker hyperfine interactions would then be preferable.

An important conclusion of the above discussion is that, in the optimization of molecular spin qubits, it is not just the value of $T_{2}$ that matters but, rather, the product $2 m T_{2}$, where $m$ is the spin projection of the (tunnel-split) ground state. Using the results of the above calculations, an approximate quantitative criterion can be derived. A molecular qubit candidate must fulfill $2m T_{2} > 70 \mu$s in order to be potentially useful for this application.

\subsection{Tuning the spin qubits}
\label{Tuning}
\begin{figure}[htb]
\centering
  \includegraphics[width=0.4\textwidth]{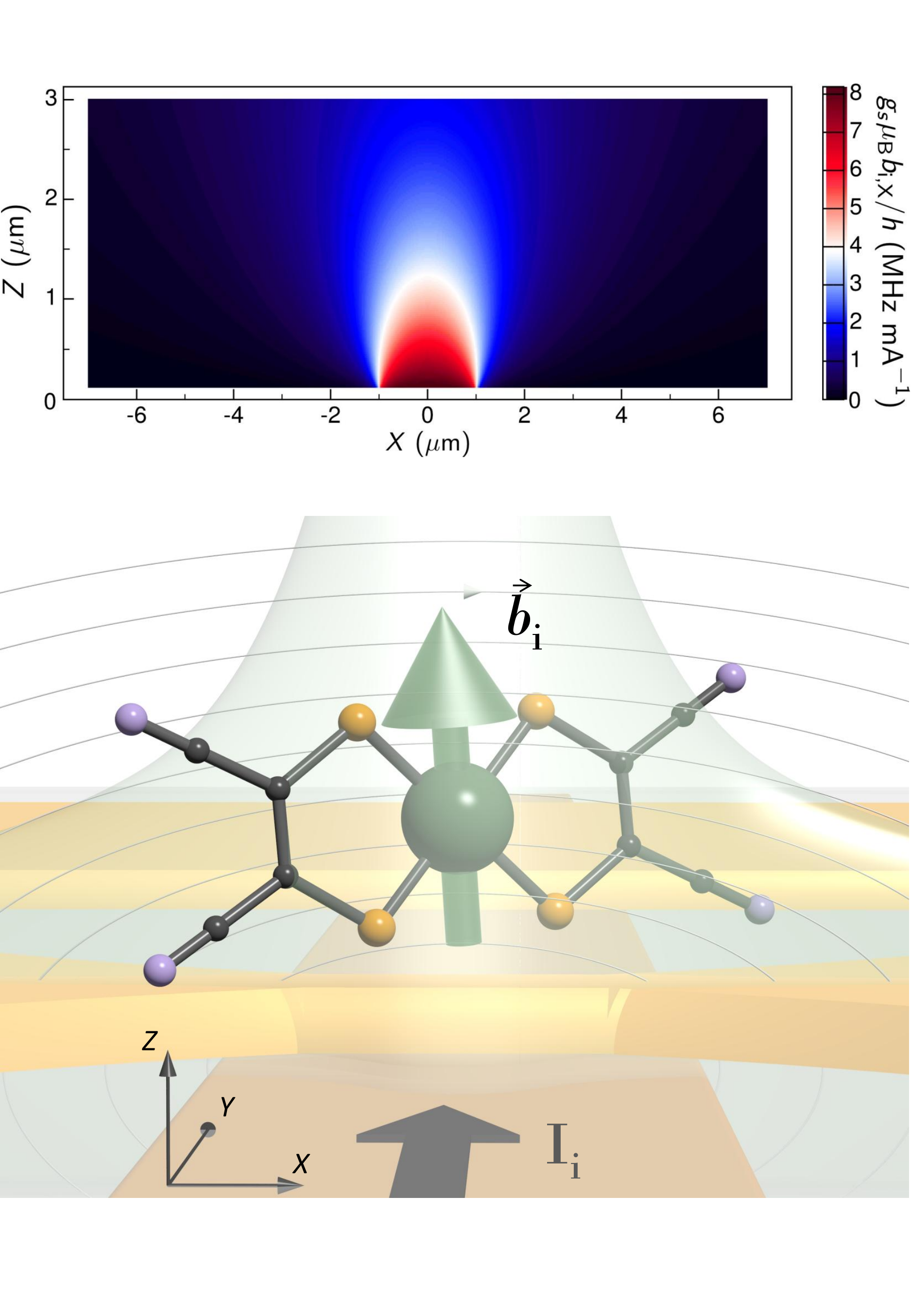}
  \caption{Top: Energy tuning of a spin qubit, generated by a current flowing through a $100$ nm thick, $2$ microns wide superconducting line located $100$ nm under the resonator plane, calculated as a function of the location of the molecule. Bottom: Artistic image of the device in the close neighborhood of a nanoconstricition hosting a molecular spin qubit. Here, $I_{\rm i}$ is the electrical current flowing via the auxiliary line and $\vec{b}_{\rm i}$ is the magnetic field the this current generates.}
  \label{fgr:Tuning}
\end{figure}

Also relevant for this proposal is $\Delta_{\rm i}$, which measures the energy detuning of each spin qubit with respect to the photons trapped in the resonator. As a starting condition, all qubits can be taken close to resonance, {\em i.e.} $\Delta_{\rm i}\simeq 0$, using an homogeneous external magnetic field $\vec{B}$. Then, each of them can be finely tuned around this condition using the field $\vec{b}_{\rm{dc,i}}$ generated by the auxiliary lines n order to either read out their spin states or induce effective qubit-qubit couplings. The set-up is shown schematically in Fig. \ref{fgr:Tuning}. Arrays of equally spaced $2$ microns wide and $100$ nm thick superconducting lines can be fabricated by optical lithography and then isolated from the resonator lines by a thin ($100$ nm) insulating film. Suitable choices for the latter material can be either SiN or Al$_{2}$O$_{3}$, whose dielectric constants are close to those of silicon or sapphire that are commonly used as substrates to fabricate the chips. The fact that the nanoconstrictions have dimensions comparable to the superconducting penetration depth, or even smaller, largely suppresses the screening of $\vec{\rm b}_{\rm i}$ by the central line of the resonator in these regions.

The magnetic field generated by each line can be easily computed. Results of these calculations, which give the energy tuning $\Delta_{\rm i} \simeq g_{S}\mu_{\rm B}b_{\rm{dc,i}}$ as a function of the location of the molecule, are shown in Fig \ref{fgr:Tuning}. These results show that values of $\Delta_{\rm i} \sim 50$ MHz can be obtained for molecules located near the nanoconstrictions and for superconducting currents smaller than $10$ mA. These values are much larger than the resonance line widths $\omega_{\rm r}/2 \pi Q \sim 5-50$ kHz, than the spin level intrinsic line widths $\sim 1/{\rm T}_{2} \sim 10^{-3}-2$ MHz (Fig. \ref{fgr:Coherence}), and than the maximum attainable coupling strengths $g_{\rm i} \sim 0.1$ MHz (Fig. \ref{fgr:Coupling}). Therefore, using these lines it is possible to properly detune each spin from the resonator and from other qubits, as required. An additional important requirement is that the influence of neighbouring lines on a given qubit is minimized in order to avoid any cross-talk between different nodes. Using the results shown in Fig. \ref{fgr:Tuning}, we find hat the different field components generated by one of these lines decay by more than two (out-of plane) or three (in plane) orders of  magnitude for separations larger than $10$ microns. If required, their mutual influence can be reduced by inserting additional ground lines in between the tuning lines. It can be concluded that a proper isolation can be achieved by separating nearest neighbour qubits by a distance of at least \SI{10}{\micro\meter}.

The same auxiliary wave guides can also be used to apply ac magnetic field pulses $\vec{b}_{\rm{ac,i}}\cos(\omega t)$ which induce single qubit operations. Since the frequencies of these coherent spin rotations are also determined by the magnetic field amplitude $\vec{b}_{\rm{ac,i}}$, operation frequencies faster than $10-50$ MHz can be attained in this manner.

\section{Integration of molecular spin qubits into superconducting circuits}
\label{Integration}
In this architecture, each constriction is coupled to only one molecule. This is probably one of the most challenging aspects of the proposal. Why it is a necessary condition can be easily understood. The proper definition, read-out and coherent control of each spin qubit is based on the fact that only one transition between two spin states is resonant with the photons. Clearly, this condition breaks down for an ensemble of identical, noninteracting molecular spins, for which degeneracies exist between different such transitions.\cite{Wesenberg2009,Imamoglu2009} However, this condition also ensures that we  profit the most from the great potential of molecular systems for attaining very large quantum information densities and from their design versatility and that spins are protected from dipole-dipole interactions. These aspects will be considered in the next section. Here, we discuss possible strategies to properly integrate molecular spin qubits into the devices.

Even though the goal is to have only one molecule contributing to the coupling at each site ${\rm i}$, the integration itself could be done by either transferring molecules in solution or molecules forming small pre-defined frameworks. However, it is then necessary to ensure that only one molecule from the deposit has a non-vanishing coupling to the resonator. This condition can be met provided that the starting material (either the solution or the framework) is magnetically diluted to such a point that the probability of two spins being sufficiently close to a resonator constriction is statistically very low. This trick has been used in the coherent control and read out of individual magnetic impurities in semiconductors.\cite{Pla2012} In order to simplify the device operation, energy gaps $\hbar \omega_{\rm i}$ and spin-photon couplings $g_{\rm i}$ of different qubits must also be very close to each other, although some inhomogeneities can be compensated using the energy bias $\Delta_{\rm i}$ generated by the auxiliary lines. This requirement implies that molecules not only need to be chemically identical but also need to orient in a preferred manner.

\begin{figure}[p]
\centering
\includegraphics[width=0.7\textwidth]{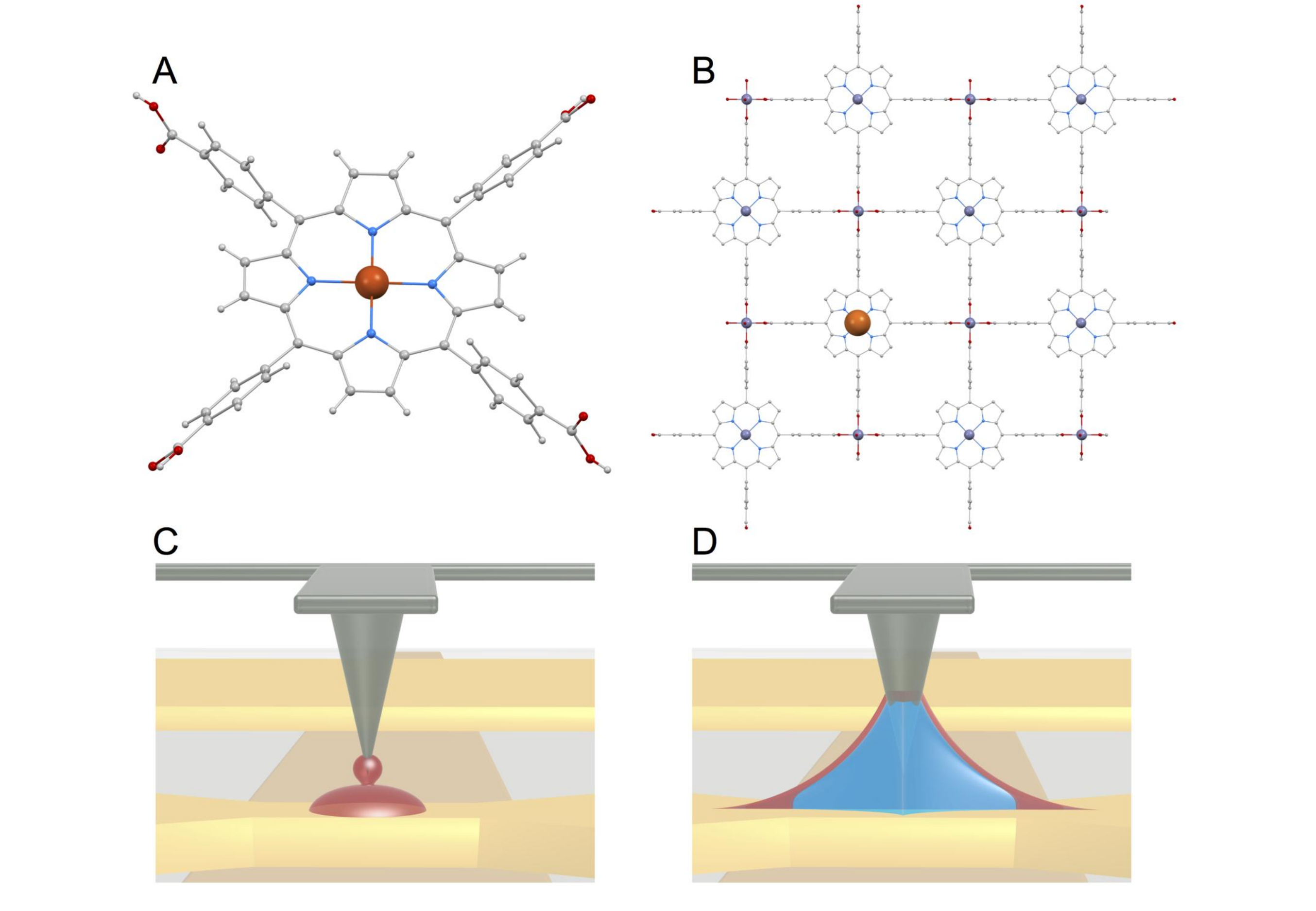}
\caption{Top. A, Molecular structure of Cu(II)tetracarboxyphenylporphyrin (CuTCPP), a candidate spin qubit that can be used for direct deposition after adequate functionalization (for example through esterification) or as a node for the formation of a $2D$ network. B, portion of a diluted $2D$ network built from a mixture of CuTCPP and of its diamagnetic analogue ZnTCPP, connected through Zn(II)$_{2}$ carboxylate paddle wheels. Colour code: dark orange, Cu(II), light violet, Zn(II), red, O, blue N, grey, C, light grey, H.
Bottom. Schematic representation of some of the envisioned strategies to integrate spin qubits into superconducting nanoconstrictions: C, chemical reactor vessel strategy in which the tip of an atomic force microscope deposits drops containing either the functionalized spin qubit molecule to react directly with the substrate or the reaction mixture of a spin qubit and a linker to form locally a $2D$ network; D, ink mixtures strategy towards the on-surface formation of a $2D$ network. An hydrophobic reagent (for example CuTCPP) remains over the meniscus surface (dark red) while an hydrophilic reagent (for example a Zn(II) salt) runs through the aqueous meniscus (blue), thereby confining on the substrate the $2D$ network formed at the interface.}
\label{fgr:Integration}
\end{figure}

Integration of spin qubits as single molecules benefits from the progress made in the last decade on the surface deposition of molecular nanomagnets.\cite{Gatteschi2009,Cornia2011,Domingo2012} Functionalization of the molecule and/or the substrate to allow specific covalent or other strong interactions between them has given access to a variety of sub-monolayer deposits of various molecular nanomagnets (mostly analogues of the prototypical [Mn$_{12}$], [Fe$_{4}$] and [TbPc$_{2}$)] species), on different substrates. In certain cases, the robustness of their quantum magnetic properties has been shown experimentally.\cite{Mannini2010} In most studies, however, the precise location of the molecules is not controlled, giving rise to a random disposition/separation on the substrate.\cite{Gatteschi2009,Cornia2011,Domingo2012} A remarkable example in this respect is the use of the strong $\pi-\pi$ interaction of a pyrene arm appended to a [TbPc$_{2}$] double-decker molecule to favor its specific binding to a carbon nanotube-based device. This allowed detecting the strong spin-phonon coupling between the molecular spin and the nanotube, which acts as a mechanical resonator.\cite{Ganzhorn2013}

Such specific interaction of a molecule with a certain area of the surface allows fixing it at the desired location, albeit it does not necessarily help controlling the number of molecules deposited in a given area. It is worth mentioning that the resonator Nb surface will be covered by a native thin layer of Nb oxide. Useful chemical functions to append the spin qubit molecules would then be chlorosilane, phosphonate or carboxylate, since they are able to efficiently bind directly to a metal oxide surface,\cite{Pujari2014} either through covalent bonds or via strong hydrogen bonds. Alternatives involve the prior removal of the thin oxide layer (e.g. by stripping with HF). Then, appending a thiol to the molecule\cite{Mannini2010} or stacking aromatic clouds of molecules such as phtalocyanine or porphyrine complexes can become useful routes to strongly bind the molecular spin qubits to the metallic surface.\cite{Katoh2010,Stepanow2010} However, these direct surface depositions should be localized onto the nanoconstrictions and therefore, they have to rely on lithographic methods, since deposition on other areas of the device with similar reactivities (rest of the resonator line, neighbouring lines, the sapphire substrate or the alumina or SiN insulating layer) has to be avoided. Dip-Pen Nanolithography (DPN) has already been used to deliver small droplets of magnetic molecules onto specific areas of superconducting sensors.\cite{Martinez-Perez2011,Bellido2013} Another approach could involve localized pre-functionalization of the constriction, entailing a different reactivity to the area of interest and therefore allowing the specific attachment of molecules with an adequate function. Here, DPN can also be used to form a self-assembled monolayer (SAM) on a specific area of the Nb oxide surface using concentrated droplets of either a phosphonate or a chlorosilane bearing the chemical function that will bind the spin qubit molecule. This strategy however implies that the molecule would be located at a distance from the surface. Clearly, because the constrictions area is rather large with respect to the size of the molecule, very dilute solutions will have to be used to limit the number of molecules deposited. Among the synthetic systems for which a reasonable spin coherence has been demonstrated, obvious candidates suitable for such direct surface anchoring would be:

\begin{enumerate}
\item Cu(II) and V(IV)O phtalocyanine (Pc) molecules and by extension their porphyrin (Pp) analogues, due to; i) their versatile chemistry, allowing many substituents to be grafted on the Pc or Pp deck, ii) their electro-neutrality, iii) the likely small effect that the deck functionalization and surface deposition will have on their spin coherence times, since the rigid environment of the metal ion will remain unchanged. Similar arguments apply to other vanadyl complexes reported very recently.\cite{Sessoli2016}
\item Ln polyoxometalates such as [GdW$_{30}$], [GdW$_{10}$], or [HoW$_{10}$] due to; i) the robustness of the polyoxometalate core, ii) the availability of procedures to graft functions on the POM outer shell;\cite{Proust2008}  iii) the availability of methods to graft POMs on surfaces in an ordered manner, for which the POM typical negative charge has not been a limitation\cite{Lombana2016}
\item Heterometallic [Cr$_{7}$Ni] rings due to; i) their reported versatile coordination and supramolecular chemistry allowing their use as a building block\cite{McInnes2015}  and ii) previous studies of deposition on metallic surfaces that have shown the robustness of the molecular properties\cite{Ghirri2011,Rath2013}
\item Neutral asymmetric [LnLn$^{\prime}$]  complexes,\cite{Aguila2014}  due to i) their outer carboxylate functions that may bind to surfaces, ii) their (relative) stability in solution and preliminary evidence for DPN deposition  iii) the potential to implement more than one qubit and iv) their adjustable Ln/Ln$^{\prime}$ composition (see next section).
\end{enumerate}

Unfortunately, the charged nature of the spin carriers in (PPh$_{4}$)$_{2}$ [Cu(mnt)$_{2}$]or in (PPh$_{4}$)$_{2}$[V(C$_{8}$S$_{8}$)$_{2}$], makes them less appealing candidates, even though they exhibit the longest coherence times measured so far.

Regarding the transfer of molecules forming small pre-defined ensembles, the required identity of all molecules and proper isolation from each other can be accessed through the periodicity provided by $2D$ networks, within which the spin qubit would be acting as node. The subjacent covalent or metal-organic framework (so-called COF and MOF respectively) will enforce the strict identity and homogeneous orientation of all molecules/nodes, while a proper adjustment of the dilution with a non-magnetic analogue node can provide the necessary control on the number of qubits per surface area. The surface-confined assembly of $2D$ architectures is actually the subject of intense research. On-surface COFs\cite{Colson2013} and MOFs\cite{Slater2011,Makiura2013,ElGarah2014} have both  been successfully formed, with a high degree of structural order up to the micrometer scale.\cite{Bartels2010} The former materials provide higher thermal and chemical stability, but in general do not guarantee error correction during the assembly due to the irreversible formation of covalent bonds. The latter systems may allow adaptation of the $2D$ network to the surface defects, as shown using flexible linkers.\cite{Kley2012} Importantly, both types of $2D$ domains can in principle be formed locally through either one or several of the following lithographic strategies, in general after the formation of an adequate SAM: i) patterning droplets containing the spin qubit building block and linker, thus confining the reaction within the deposited volume, possibly after thermal activation; ii) use of microfluidic pens to deliver small volumes of precursors at specific locations of the surface and perform the reaction locally;\cite{Carbonell2013}  iii) confined in-plane deposition induced by the use of inks mixtures with different solubility;\cite{Salaita2005}  iv) in-plane deposition through the receding meniscus technique, i.e. controlling the relative contribution of evaporation and viscous forces, forcing the system to work into the liquid viscosity driving deposition.\cite{LeBerre2009}

For the elaboration of such surface-induced frameworks, a few synthetic systems appear as potentially good nodes, for which non-magnetic analogues are available:

1)	Cu(II) and V(IV)O tetrasubstituted porphyrins (for example CuTCPP, see Fig. \ref{fgr:Integration}; diamagnetic analogues can be with either Zn(II), Ti(IV)O or Ni(II)) due to: i) the existence of a number of $2D$ and $3D$ MOFs and COFs based on these or similar molecules,\cite{Burnett2013, Lin2016}  ii) the fact that ordered $2D$ networks have been deposited successfully on surfaces;\cite{Makiura2010}  iii) their versatile chemistry and relative ease of purification, which should allow the modulation of the $2D$ framework\cite{Makiura2014}

2)	heterometallic [Cr$_{7}$Ni] rings (diamagnetic analogue could be the [Cr$_{8}$] ring due to its singlet ground state), given the existence of some extended networks built on these building blocks and their versatile chemistry;\cite{McInnes2015} by extension, any spin qubit molecule with exchangeable carboxylates or other labile coordination sites, such as triangular [M$_{3}$] complexes.\cite{Mitrikas2008,Walsh2015}

At this stage, it is still unclear which strategy will prove more effective. We are currently exploring several of them, mostly using porphyrin synthetic systems.

\section{Potential for scalability}
\label{Scalability}
In some of the previous sections, we have often used the analogy of the present proposal with similar schemes that make use of superconducting circuits, like transmons, to realize qubits. Considering the underlying physics, both schemes are similar. However, their parameters differ. Whereas superconducting circuits strongly couple to the electrical rf field generated by the resonators, attaining this limit for a single molecular spin is very challenging. In return, molecular spins have properties that make them very attractive for building dense and complex quantum computational architectures.

The first and obvious one, which they share with other microscopic qubits like impurity spins in semiconductors, is the fact of being very small, with lateral dimensions of about $1$ nm, thus much smaller than solid state qubits. As mentioned above, this fact allows enhancing the coupling to photons near narrow areas at the edges of the superconducting wires and in nanoconstrictions. The operational architecture needed to control and read out each qubit occupies just a few microns wide area and is separated by about \SI{10}{\micro\meter} from its nearest neighbours. By contrast, the region in which the microwave magnetic field $b_{\rm r}$ generated by the resonator stays close to its maximum scales with the wavelength of microwave photons, between $66$ mm for $\omega_{\rm r}/2 \pi = 1$ GHz and $6.6$ mm for $\omega_{\rm r}/2 \pi = 10$ GHz, and it is therefore much wider. One can then see from these considerations that a single chip can host, and couple to, a very large number $N > 100$ of qubits. The limit in the density of quantum information processable by each device would probably be set by the influence that the presence of  nanoconstrictions and auxiliary lines has on the circuit losses, which will eventually limit the attainment of the strong coupling condition $g_{\rm i} \kappa > 1$. Also, reading out $N$ qubits in a single transmission experiment requires that the resonance frequencies that correspond to each logical state of the array (say $1 0 0 1 \cdots 0 0 1$) are different. This can be achieved by making $\Delta_{\rm i}$ of all spins different from each other. Besides, these frequencies must also be separated by shifts larger than the resonance width $\kappa$. This second requirement imposes that $g_{\rm i} > N \kappa$, thus going beyond the standard strong coupling regime by a factor $N$.

But molecules are not just "small", but also very reproducible and flexible objects. In contrast with "natural" magnetic defects, such as NV$^{-}$ centers in diamond\cite{Jelezko2004} or P impurities in silicon,\cite{Pla2012} magnetic molecules are artificial objects synthesized by chemical methods. One of the advantages, which has been discussed in the previous section, is that molecules are often stable in solution. This considerably eases the preparation of different material forms and, what is essential for the present purposes, their integration into devices.

Chemical design offers also nearly unbound possibilities to modify the properties of the magnetic core. In particular, each molecule can host and stabilize not just one, but several addressable qubits. We recently reviewed the potential and first results of using coordination complexes to host $2$-qubit quantum gates.\cite{Aromi2012} Possible strategies include: a) the elaboration of molecules containing two well-defined paramagnetic metal ion clusters, each acting as single spin qubit and weakly coupled to the other one, and b) the design of dinuclear complexes of anisotropic metal ions, specifically lanthanides, possessing dissimilar environments and a weak exchange interaction. Since then, exciting results showing the validity of both approaches have been reported. A [Tb]$_{2}$ and a [CeEr] complexes were shown to fulfil all requisites to embody universal C-NOT quantum gates.\cite{Luis2011,Aguila2014} Spin coherence times of a molecular $2-$qubit gate were also measured for the first time on the latter complex. Although ${\rm T}_{2}$ is still relatively short ($\simeq 410$ ns) these experiments show that coherent manipulations of these systems are nevertheless feasible.  Even more recently, a family of [Cr$_{7}$Ni] dimers with a variety of linking groups has been studied and realizations of C-NOT and C-PHASE gates based on these supramolecular systems have been proposed.\cite{Ferrando-Soria2016,Fernandez2016} The additional spin degrees of freedom introduce a kind of extra dimension to the Hilbert space along which computation can be scaled up. However, perhaps the most interesting application of such extra states is the development of on-site protocols to protect qubits from decoherence. For this, it is not even necessary that the number of spin states be a multiple of $2$. Embedding a qubit in a system with a Hilbert space of dimension $d>2$ (a "qudit") enables correcting some specific errors.\cite{Pirandola2008}

The operations required to control the molecular gates or the qudits are combinations of phase and energy shifts, which can be induced by dc field pulses $b_{\rm{i, dc}}$, and of resonant transitions between different levels, induced by ac pulses $b_{\rm{i, ac}}$. In connection with the present proposal, an important limitation is that, in order to be accessible, all spin energy levels must be separated by gaps comparable to $\omega$, which as said above lies between $1$ and $10$ GHz. In addition, these energy gaps must all be different from each other, in order to be addressable (e.g. by varying $\omega$), but not too different. The latter requirement ensures that different transitions can also be tuned with respect to the fixed resonator frequency $\omega_{\rm r}$ using the energy bias $\Delta_{\rm i} \sim 5-50$ MHz that can be generated by the auxiliary lines. This condition seems to be fulfilled by molecular gates made of true $S=1/2$ qubits. In the case of molecules made of lanthanide ions, it would be necessary to look for those having the smallest possible magnetic anisotropy, e.g. Gd(III).

\section{Summary and outlook}
\label{Summary}
In this work, we have put forward a first proposal for a scalable magnetic quantum processor involving individual molecular spin qubits coupled to superconducting resonators and to superconducting open lines. This hybrid device allows performing basic operations on each individual qubit as well as switching on and off the effective couplings between any two qubits that are required to perform two-qubit gates. Thanks to the microscopic size, identical nature and design versatility of the molecular qubits, this architecture would enable processing high quantum information densities, unparalleled by other existing solid-state platforms. Besides, calculations show that the proposal is feasible, although very challenging.

Some of these challenges set specific targets for the development of suitable molecules and new methods to manipulate them. A crucial milestone in this endeavor is to attain the coherent or strong coupling regime, that is, to make the coupling strength $g_{\rm i}$ of individual molecular spins to single photons trapped in the resonator sufficiently large as compared to the dissipation rates of both the spins $T_{2}^{-1}$ and the superconducting circuit $\kappa$. In order to reach this limit the magnetic field generated by the resonator needs to be enhanced locally by reducing the diameter of its central superconducting line to values of order of a few tens of nm. In addition, spin coherence times need to be improved to the limit. However important, enhancing $T_{2}$ (and $T_{1}$) is not all that is necessary. For the case of $S=1/2$ molecular complexes, $T_{2}$ values close to a ms are necessary to compensate for their relatively weak coupling. Yet, in this case the decoherence time of the circuit might become the limiting factor. Stronger couplings can be attained with qubits having $S>1/2$. A promising strategy is the use of clock transitions between high-spin states of lanthanide ions. In this case, the strong coupling could be reached provided that $T_{2}$ is enhanced to values of more than $10-50 \mu$s. An alternative would be to develop qubit candidates that couple to the electric field of the photons, e.g. via the modulation of the crystal field and the spin-orbit interaction. Perhaps the most difficult challenge is related to the need of properly integrating the molecular qubits into specific areas of the circuit, namely, on the nanoconstrictions and close to the auxiliary superconducting lines that tune their energies and induce single qubit operations. Also in this aspect, it will be necessary to go beyond the limits of present technologies. Potentially promising strategies combine chemical functionalization with nanolithography methods. Finally, this proposal underlines the need to characterize spin relaxation and decoherence of isolated spins grafted onto superconducting substrates.

\section{Acknowledgements}
\label{Thanks}
The authors acknowledge funding from the Spanish Ministry of Economy and Competitivity (MINECO) through grants FIS2014-55867-P, MAT2014-53432-C5-1-R, MAT2014-53961-R, MAT2015-68204-R, and MAT2015-70868-ERC, from the European Research Council through grant ERC-2010-StG (258060 FuncMolQIP) and from a TOP grant of the Technical University of Vienna.

%%%END OF MAIN TEXT%%%


%\balance


%%%REFERENCES%%%
\bibliographystyle{h-physrev3} %the RSC's .bst file
\bibliography{Jenkins_resub} %You need to replace "rsc" on this line with the name of your .bib file

\end{document}